\documentclass[onecolumn,amsmath,amssymb]{qm2008_abs}
\usepackage{graphicx}% Include figure files
\usepackage{dcolumn}% Align table columns on decimal point
\usepackage{bm}% bold math
\usepackage{wrapfig} 
\usepackage{float}
\begin{document}

\title{{\Large $\psi\prime$ to $J/\psi$ Ratio Measurements in PHENIX at RHIC}}% Force line breaks with \\

\bigskip
\author{\large Maris\'ilvia Donadelli for the PHENIX Collaboration }
\email{mari@fig.if.usp.br}
\affiliation{University of S\~ao Paulo, S\~ao Paulo, Brazil}
\bigskip

\begin{abstract}
\leftskip1.0cm
\rightskip1.0cm

The ratio of the $\psi'$ over the $J/\psi$ production cross section in the dielectron channel has been measured in  $\sqrt{s}=$ 200 GeV $p+p$ collisions with the PHENIX detector at RHIC. The analysis is based on fitting of the dielectron invariant mass spectra in the area around the $J/\psi$ and $\psi'$ signals in order to extract a ratio $\psi'$ over  $J/\psi$ of 0.019$\pm$0.005$($stat$)\pm$0.002$($sys$)$ and a fractional feed-down contribution to $J/\psi$ from $\psi^\prime$ of $8.6 \pm 2.5 \%$.

\end{abstract}

\maketitle

\section{Introduction}

$J/\psi$ production in hadronic collisions can occur in part through the production of higher excited resonances, $\psi^\prime$, $\chi_{c}$, which decay into the ground state. For $J/\psi$ it is known experimentally that about 40\% of  hadroproduction rate comes from feed-down of higher mass resonances \cite{ISR_CERN}, \cite{SPS_CERN}, \cite{Fermilab}. Measurement of excited charmonium states is of key importance to account for the feed-down contributions. Thus, this is an aditional test for different production mechanisms of charmonium.

\section{Measurement method}

 The number of events in the $J/\psi$ peak, $N_{J/\psi}$, is given by:
\begin{eqnarray}
N_{J/\psi}=\sigma(J/\psi) \cdot \mathcal{B}(J/\psi \rightarrow l^{+}l^{-}) \cdot \mathcal{L}\cdot \epsilon,
\label{fig:eventsjpsi}
\end{eqnarray}

i.e. the product of the cross section $(\sigma(J/\psi))$, the branching ratio into dilepton pairs $(\mathcal{B})$, the integrated luminosity $\mathcal{L}$ and the total reconstruction and trigger efficiencies,
as well as acceptance $(\epsilon)$. 
The luminosity is identical for all charmonium states, and therefore cancels in the ratios. The ratio of the $\psi'$ and  $J/\psi$ cross sections in the $e^{+}e^{-}$ channel, $R_{\psi'}(e)$ is equal to:

\begin{eqnarray}
R_{\psi'}(e)= \frac{\mathcal{B}'\cdot \sigma'}{\mathcal{B}\cdot \sigma}= \frac{N_{\psi'}}{N_{J/\psi}} \cdot \frac {\epsilon}{\epsilon'}
\label{fig:ratio}
\end{eqnarray}

where $e$ denotes the leptonic decay channel, $\sigma(\sigma')$ is the $J/\psi(\psi')$ production cross section and $\mathcal{B}(\mathcal{B'})$ is the branching ratio for the $e^{+}e^{-}$ decay into the $J/\psi (\psi')$ meson.

The ratio $\frac{N_{\psi'}}{N_{J/\psi}}$ or 'raw $\frac{\psi'}{J/\psi}$ ratio' is defined from the fit of the $J/\psi$ and $\psi'$ signals and must be corrected by the efficiency ratio, $\frac{\epsilon}{\epsilon'}$, where $\epsilon$ is defined above and $\epsilon'$ is the corresponding efficiency for detection of $\psi'$ mesons. The efficiencies are evaluated by simulation as described in \ref{analysis}.

\section{\label{setup}Experimental Setup}

Data collected during the 2006 RHIC run  with the PHENIX central arms spectrometers which cover $|\eta| <$ 0.35 in pseudorapidity  and 2 $\times \pi/$2 in azimuth were used for this letter. Electrons and positrons are reconstructed in the central arms \cite{PHENIX} using Drift Chambers (DC) and Pad Chambers (PC), located outside an axial magnetic field, and are identified  by hits in the Ring Imaging Cherenkov detector (RICH) and by matching the  momentum with the energy measured in an Electromagnetic Calorimeter (EmCal).

This analysis is based on collision events triggered by minimum activity in the Beam Beam Counters at 3.0 $<|\eta|<$ 3.9 (minimum bias trigger), and associated hits between the RICH and a 400 MeV cluster in the EmCal. The data sample cosisted of 88 billion minimum bias events.   

\section{\label{analysis}Analysis and discussion}

Two sets of electron identification cuts were used to count $J/\psi$ and $\psi^\prime$. A first set, named 'tight' with a match between energy and momentum satisfying $(E/p-1) \geq$ -3 standard deviations $(\sigma)$, interaction-vertex cut of $\pm$30 cm. The second set, named 'loose' with energy-momentum requirement (E/p-1) $\geq$ -4$\sigma$, interaction-vertex cut of $\pm$30 cm and a minimum number of DC hits used to reconstruct each track of the electron-positron pair.

The number of $J/\psi$ candidates is obtained by counting the unlike sign pairs in the mass window (2.88-3.32) GeV/c$^{2}$ and the $\psi^\prime$ candidates in the mass window (3.48-3.92) GeV/c$^{2}$. The combinatorial background subtraction is performed as the sum of the like sign pairs.  The subtracted plot contains charmonium resonances and physical background made of correlated $(e^+e^-)$ pairs from  Drell Yan  and leptonic decays of charm and bottom. From PYTHIA the $(e^+e^-)$ invariant mass distribution was obtained for all sources of physical background under the $J/\psi$ and $\psi^\prime$ as shown in Figure 1. The resulting sum of all the three components was fit with a power law. 

To count the number of $J/\psi$s and $\psi^\prime$s  different fit procedures were applied for each one of the electron identification cuts. Figure 2 shows a double gaussian shape for the $J/\psi$ peak, a single gaussian for $\psi^\prime$. The background was fit with a power law, the $J/\psi$-$\psi^\prime$ mass difference was fixed with the Particle Data Group value (0.589 GeV/$c^{2}$) and both $J/\psi$ and $\psi^\prime$ widths varied independently. For this fit, loose electron identification cuts  set was applied.

\begin{figure}[h]
 \begin{minipage}[b]{.48\linewidth}
\begin{center}
\includegraphics[
  width=0.8\columnwidth]{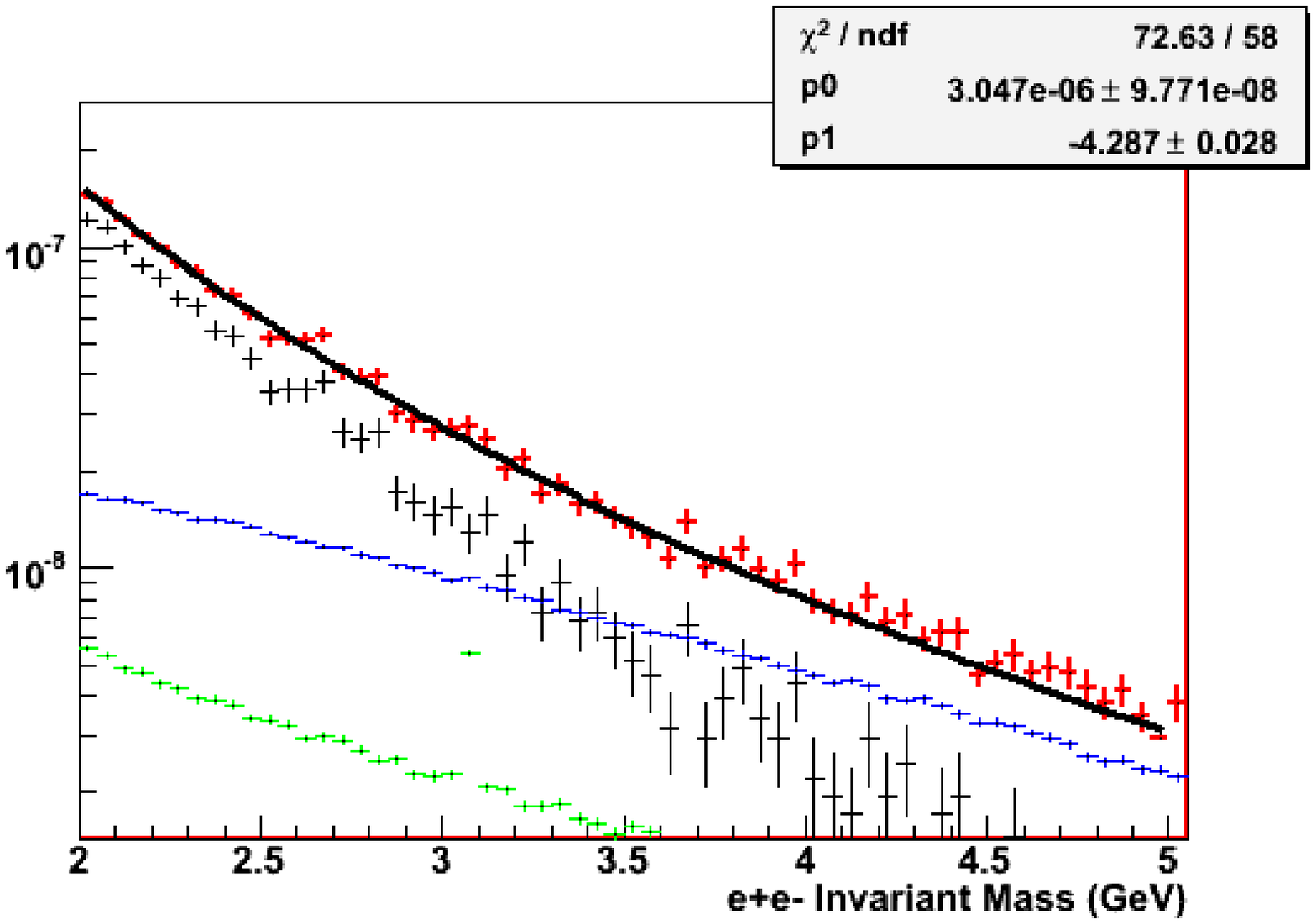}
\end{center}
\vspace{-20pt}
\label{fig:fit-power}
\caption{$e^{+}e^{-}$ invariant mass distribution for all sources of physical background under $J/\psi$ and $\psi^\prime$.}
\end{minipage}\hfill
 \begin{minipage}[b]{.48\linewidth}
  \begin{center}
\includegraphics[
 width=0.85\columnwidth]{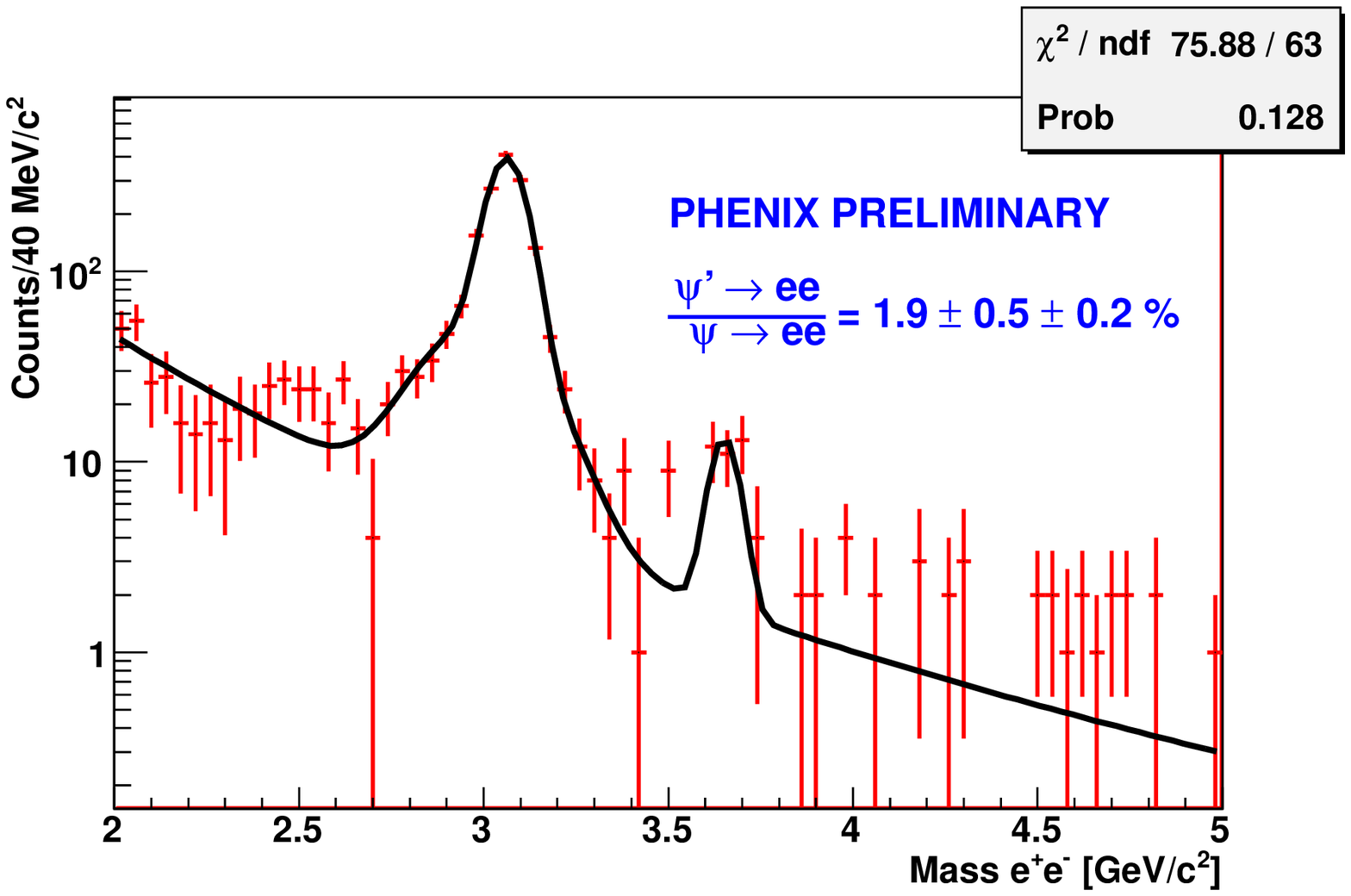}
\end{center}
\vspace{-20pt}
\label{fig:fitpreliminaryC3}
\caption{Invariant mass distribution $e^{+}e^{-}$ in the region of the $J/\psi$ and $\psi^{\prime}$ peaks for triggered dielectron events.}
\end{minipage}
\end{figure}

Acceptance and reconstruction efficiency were calculated by generating $J/\psi$ and $\psi^\prime$ with PYTHIA event generator and checking the response of PHENIX detector with a GEANT based Monte Carlo. In order to generate $J/\psi$, all  $J/\psi$ decays were turned off except the dielectron one. To account for 
direct and indirect production (through $\chi$ and $\psi^\prime$ decays) the 
subprocesses reported on the left of Table I were used. In order to generate $\psi^\prime$  particle id=100443 was requested, instead of 443 and
all  $\psi^\prime$ decay modes were turned  off except the dielectron one plus the following on the right of Table I.

\begin{table}
\begin{minipage}[b]{.48\linewidth}
\begin{tabular}{|c|c|}
    \hline\hline
96 &  Semihard QCD 2 $\rightarrow $2 \\
86 & $ g + g \rightarrow J/\Psi + g$ \\
87 & $ g + g \rightarrow \chi_{0c} + g$ \\
88 & $ g + g \rightarrow \chi_{1c} + g$  \\
89 & $ g + g \rightarrow \chi_{2c} + g$ \\
104 & $  g + g \rightarrow \chi_{0c}$ \\
105 & $ g + g \rightarrow \chi_{2c} $ \\
106 & $ g + g \rightarrow J/\Psi + \gamma $\\
    \hline\hline
\end{tabular}
\end{minipage}[b]\hfill
\begin{minipage}[b]{.48\linewidth}
\begin{tabular}{|c|c|}
    \hline\hline
96 &  Semihard QCD 2 $\rightarrow $2 \\
86 & $ g + g \rightarrow \psi\prime + g$ \\
106 & $ g + g \rightarrow \psi\prime + \gamma $\\
    \hline\hline
\end{tabular}
\end{minipage}
\caption{$J/\psi$ and $\psi^\prime$ subprocesses in PYTHIA event generator.}
\end{table}

The average transverse momentum is about 1.7~GeV for $J/\psi$ and 1.9~GeV for $\psi^\prime$ as shown in Figures 3 and 4. The electron average transverse momentum is about 1.5~GeV for electrons coming from $J/\psi$ decays and 1.8~GeV for electrons coming $\psi^\prime$.

\begin{figure}
 \begin{minipage}[H]{.48\linewidth}
\begin{center}
\includegraphics[
  width=0.85\columnwidth]{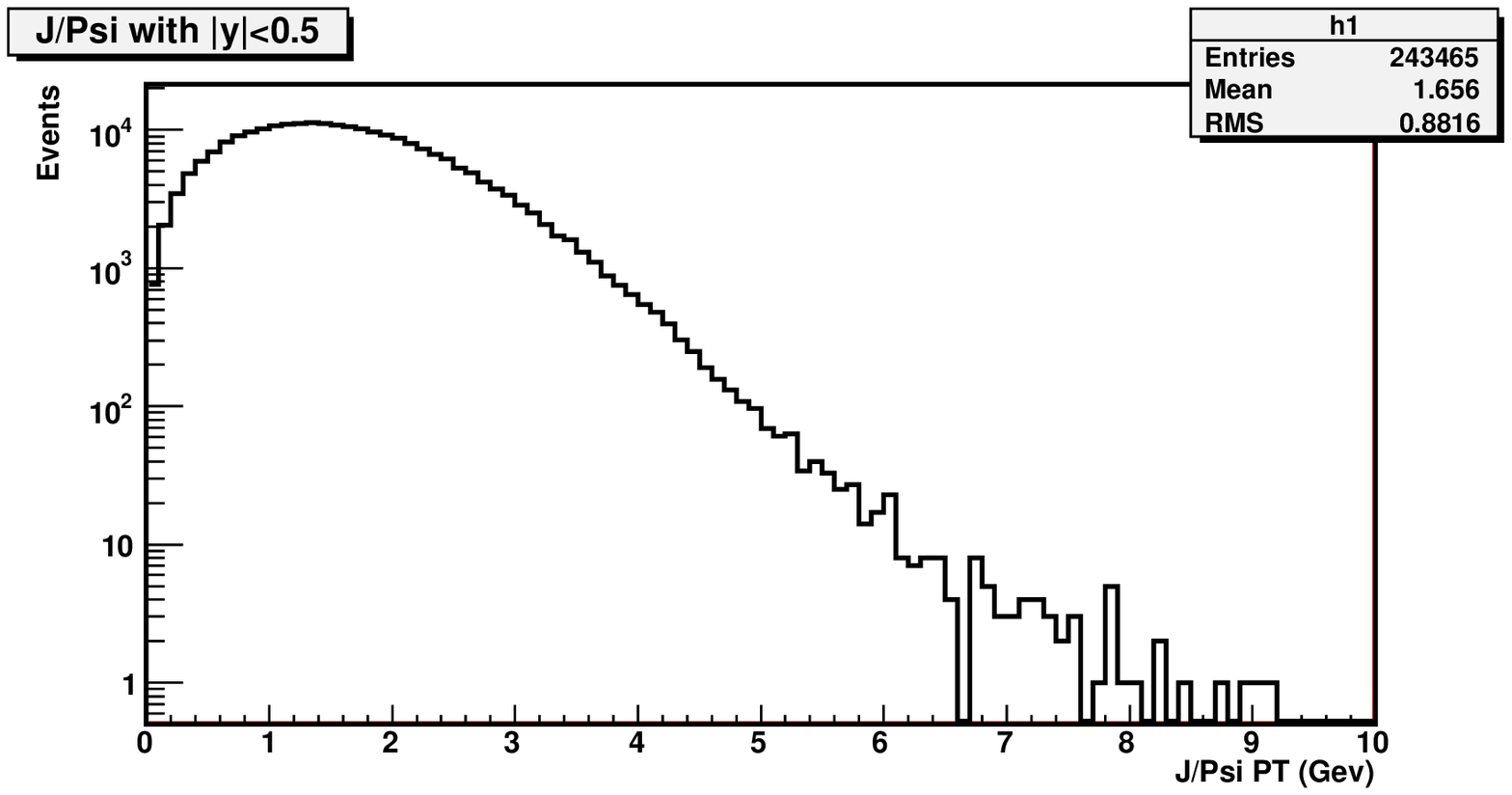}
\end{center}
\vspace{-20pt}
\label{fig:pythiajpsipt}
\caption{Transverse momentum distribution for all $J/\psi$ with rapidity $-0.5<y<0.5$ generated with PYTHIA event generator.} 
\end{minipage} \hfill
 \begin{minipage}[H]{.48\linewidth}
  \begin{center}
\includegraphics[
  width=0.85\columnwidth]{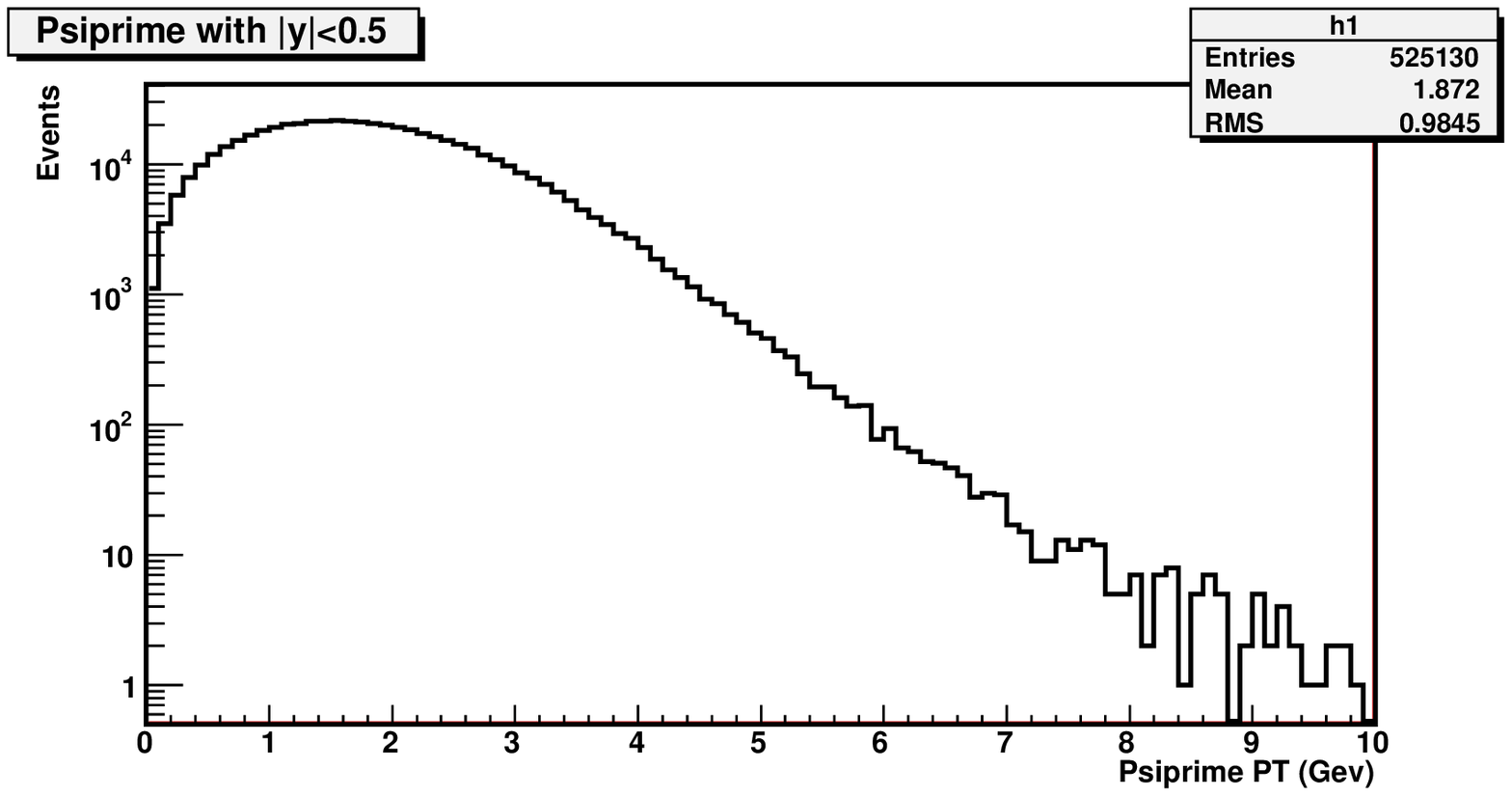}
\end{center}
\vspace{-20pt}
\label{fig:pythiapsippt}
\caption{Transverse momentum distribution for all $\psi^\prime$ with rapidity $-0.5<y<0.5$ generated with PYTHIA event generator.}

 \end{minipage}
\end{figure}

$J/\psi$ and $\psi^\prime$ with rapidity $-0.5<y<0.5$ were selected 
and  the two electrons from their decay were put through the PHENIX simulation
chain (PISA and PISAtoDST). After that, 
the reconstructed dielectron invariant mass was determined and the electron identification cuts used in the data analysis were applied to evaluate the acceptance 
and reconstruction efficiency. The {acceptance $\times$ reconstruction efficiency} for $J/\psi$ and $\psi^\prime$ are shown 
in Figures 5 and 6 respectively, both for loose electron identification cuts. The integrated acceptance $\times$ reconstruction efficiency is 2.48 for $\psi^\prime$ and 2.41 for $J/\psi$.

\begin{figure}
 \begin{minipage}[H]{.48\linewidth}
  \begin{center}
\includegraphics[
  width=0.85\columnwidth]{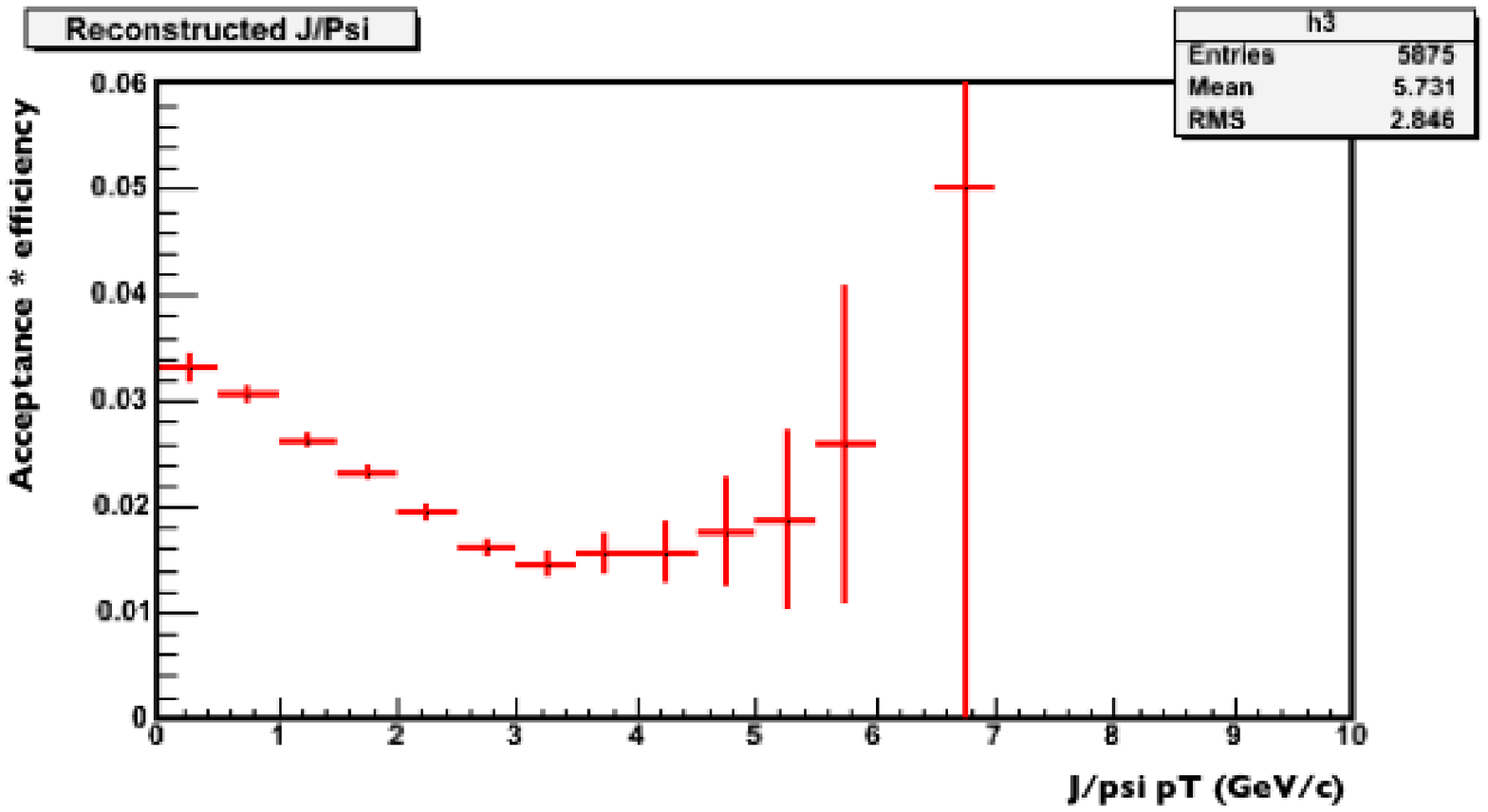}
\end{center}
\vspace{-20pt}
\label{fig:pythiajpsireceff}
\caption{The {acceptance $\times$ reconstruction efficiency} for $J/\psi$ for loose electron id cuts.}
\end{minipage} \hfill
 \begin{minipage}[H]{.48\linewidth}
  \begin{center}
\includegraphics[
  width=0.85\columnwidth]{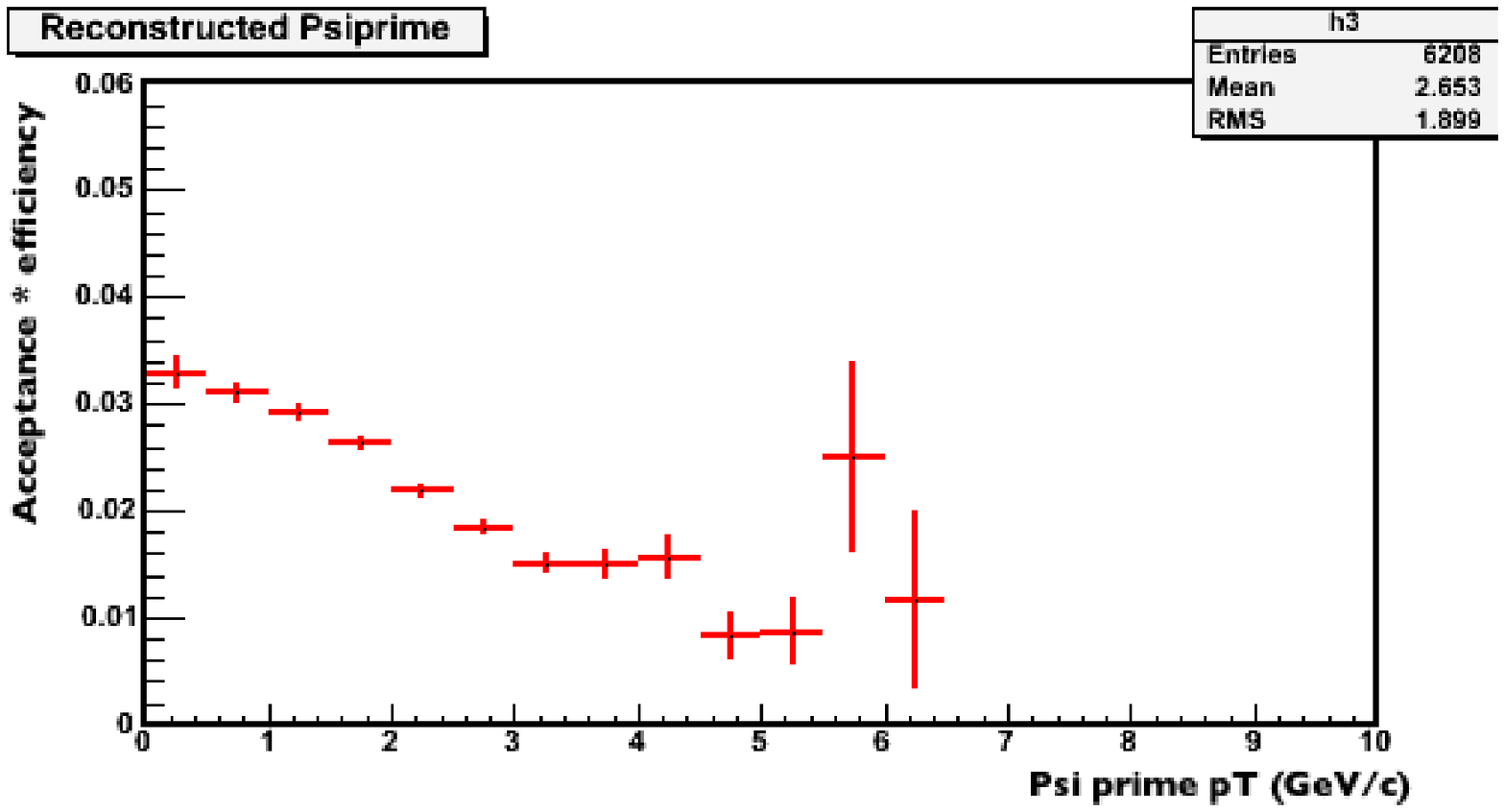}
\end{center}
\vspace{-20pt}
\label{fig:pythiapsipreceff}
\caption{The {acceptance $\times$ reconstruction efficiency} for $\psi^\prime$ for loose electron id cuts.}
 \end{minipage}
\end{figure}

Integrating over the momentum distribution predicted by PYTHIA we find that the ratio of the $\psi^\prime$ acceptance 
$\times$ reconstruction efficiency to the $J/\psi$ is $\epsilon '/\epsilon=1.03\pm0.01$.

\section{Summary}

\begin{wrapfigure}{r}{0.5\textwidth}
\vspace{-30pt}
  \begin{center}
    \includegraphics[width=0.48\textwidth]{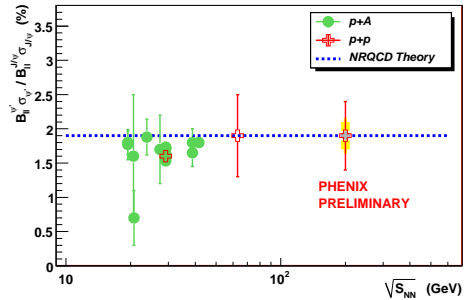}
  \end{center}
  \vspace{-20pt}
  \caption{$\psi^\prime$ over $J/\psi$ ratio as a function of the energy in the center of mass frame measured in various experiments.}
\end{wrapfigure}

The $\psi'$ over $J/\psi$ ratio is quoted as obtained with loose electron identification cuts and the fit double gaussian (for $J/\psi$), single gaussian (for $\psi^\prime$) and  power law (for the physical background) $\cdot \epsilon '/\epsilon$ (acceptance $\times$ reconstruction efficiency): 0.019 $\pm$ 0.005. The systematic error derived by taking the maximum-minimum interval for the ensemble of the various fit results and divided by $\sqrt{12}$ is $\pm0.002$. Figure 9 shows the ratio results of the experiments listed on Table II, and also the result quoted in this letter. Based on the results here presented, the fractional feed-down contribution to $J/\psi$ from $\psi^\prime$ is $8.6 \pm 2.5 \%$ in good agreement with the calculation provided by \cite{Digal}, which is $8 \pm 2\%$.

\begin{table}[H]
{\tiny
\begin{center}
\begin{tabular}{|c|c|c|c|}
    \hline\hline
\textbf{Experiment} & \textbf{target} & \textbf{Energy GeV} & \textbf{Result}\\ \hline
NA51 \cite{Abreu1} & p & 29.1&1.6$\pm$0.04\\
ISR \cite{Clark}& p&63&1.9$\pm$0.6 \\
E288 \cite{Snyder}& Be& 27.4&1.7$\pm$0.5\\
E331 \cite{Branson}&  C&20.6&0.7$\pm$0.4\\
E444 \cite{Anderson}&  C&20.6&1.6$\pm$0.09\\
E705 \cite{Antoniazzi}&  Li&23.8&1.88$\pm$0.26\\
E771 \cite{Alexopoulos}&Si&38.8&1.65$\pm$0.2\\
E789 \cite{Schub}&Au&38.8&1.8$\pm$0.2\\
NA38 \cite{Abreu2}& C&29.1&1.9$\pm$0.13\\   
NA38 \cite{Abreu2}& Al&29.1&1.36$\pm$0.35\\
NA38 \cite{Abreu2}& Cu&29.1&1.74$\pm$0.11\\
NA38 \cite{Abreu2}& W&29.1& 1.59$\pm$0.13\\
NA51 \cite{Abreu2}& d&29.1&1.71$\pm$0.04\\
NA50 \cite{Alessandro}&Be&29.1&1.73$\pm$0.04\\
NA50 \cite{Alessandro}&Al&29.1&1.73$\pm$0.05\\
NA50 \cite{Alessandro}&Cu&29.1&1.64$\pm$0.03\\
NA50 \cite{Alessandro}&Cu&29.1&1.57$\pm$0.03\\
NA50 \cite{Alessandro}&Cu&29.1&1.53$\pm$0.04\\
NA38 \cite{Lourenco}&W&19.4&1.8$\pm$0.17\\
NA38 \cite{Lourenco}&U&19.4&1.77$\pm$0.22\\
HERA-B \cite{Abt}&various&41.6&1.80$\pm$0.08\\
\hline\hline
\end{tabular}
\end{center}}
\caption{$\psi^\prime$ over $J/\psi$ ratio for different energy regimes and collision species.}
\end{table}

\end{document}